\documentclass{gGAF2e}

\usepackage{color}
\usepackage{mathrsfs}

\begin{document}

\jvol{00} \jnum{00} \jyear{2012} %\jmonth{February}

\markboth{L.~Oruba}{Geophysical and Astrophysical Fluid Dynamics}

\newcommand{\X}{\! \times \!}
%%%% User macro

\def\bmu{\bm u}
\def\bmB{\bm B}

\def\bmustar{\bm u$^\star$}

\def\bfe{\mbox{\bf e}}
\def\bfz{\mbox{\bf z}}
\def\bfr{\mbox{\bf r}}

\def\bfe{\bm e}
\def\bfz{\bm z}
\def\bfr{\bm r}

\def\dd{{\rm d}}

\def\ff{{\rm FF}}
\def\dh{{\rm DH}}
\def\bmnabla{\bm \nabla}

\newcommand*{\Ray}{{\rm Ra}}
\newcommand*{\Power}{{P}}
\newcommand*{\Pra}{{\rm Pr}}
\newcommand*{\Pm}{{\rm Pm}}
\newcommand*{\Rey}{{\rm Re}}
\newcommand*{\Rm}{{\rm Rm}}
\newcommand*{\q}{{\rm q}}
\newcommand*{\Ek}{{\rm E}}
\newcommand*{\Nu}{{\rm Nu}}
\newcommand*{\Lo}{{\rm Lo}}
\newcommand{\Ro}{{\rm Ro}}
\newcommand{\calE}{{\cal E}}
\newcommand{\fohm}{{f_{\rm ohm}}}
\newcommand{\fdip}{{f_{\rm dip}}}
\newcommand{\taudiss}{{\tau_{\rm diss}}}

\jvol{00} \jnum{00} \jyear{2012} %\jmonth{February}

\title{On the role of thermal boundary conditions in dynamo scaling laws}

\author{Ludivine Oruba $^{\ast}$
\thanks{$^\ast$Corresponding author. Email: ludivine.oruba@ens.fr}
\\\vspace{6pt}  Physics Department, Ecole Normale Sup\'{e}rieure, 24 rue Lhomond, 75005 Paris, France}
%\\
%\vspace{6pt}\received{} }

\maketitle

\begin{abstract}

In dynamo power-based scaling laws, the power $\Power$ 
injected by buoyancy forces is measured by a so-called flux-based  
Rayleigh number, denoted as $\Ray_Q^\star$ 
(see Christensen and Aubert, Geophys. J.
Int. 2006, vol. 166, pp. 97-–114).
Whereas it is widely accepted that this parameter is measured 
(as opposite to controlled) in dynamos driven by differential heating, 
the literature is much less clear concerning its nature in the case of imposed heat flux. 
We clarify this issue by highlighting that in that case, 
the $\Ray_{Q}^\star$ parameter becomes controlled only in the limit of 
large Nusselt numbers ($\Nu \gg 1$).
 
We then address the issue of the robustness of the original 
relation between $\Power$ and $\Ray_Q^\star$ with the geometry and the 
thermal boundary conditions. We show that in the cartesian geometry,
as in the spherical geometry with a central mass distribution, 
this relation is purely linear, in both differential and 
fixed-flux heating. However, we show that in the geometry commonly 
studied by geophysicists (spherical with uniform mass 
distribution), its validity places an upper-bound on 
the strength of the driving which can be envisaged in a fixed 
Ekman number simulation. An increase of the Rayleigh number 
indeed yields deviations (in terms of absolute correction) 
from the linear relation between $\Power$ and $\Ray_Q^\star$. 
We conclude that in such configurations, 
the parameter range for which $\Power$ is controlled is limited.

\end{abstract}

\section{Introduction}

Power-based scaling laws introduced in 
\cite{Chr06} have been very successful in the dynamos community, 
and have been further developed in many recent studies 
\citep[e.g.][]{JonesRev, Stelzer13, Davidson13, Oruba}. 
\cite{Davidson13} proposed a 
modified version of these original scaling laws, which is dedicated 
to planetary 
dynamos and which slightly differs from the original one because the 
non-linear inertial term is assumed to be negligible. 
\cite{Oruba} pointed out some ``weaknesses'' of the original scaling laws 
for the magnetic field strength. 
They stress that the power-based scaling laws essentially reflect 
the statistical balance between energy 
production and dissipation for saturated dynamos, and therefore work for 
any dynamo. Besides, these laws relate
measured quantities (e.g. the magnetic field strength, or the flow strength) 
to another measured quantity, which is the power injected by buoyancy forces, 
as measured by the flux-based Rayleigh number $\Ray_Q^\star$ 
\citep[see (19) in][]{Chr06}.

The flux-based Rayleigh number $\Ray_Q^\star$ involves the advected heat flow 
which is the 
difference between the time-averaged total heat flow and the conductive heat 
flow corresponding to the realized difference of temperature between both 
boundaries. There exists several contradictions and/or ambiguities in the 
literature concerning the nature of this parameter (controlled versus measured). 
Whereas there is a wide consensus on its measured nature in dynamos driven by an 
imposed difference of temperature between the boundaries, the literature is much 
less clear concerning its nature in the case of dynamos driven by an imposed 
heat flux.  
In \cite{Chr10}, it is  
suggested to be a control parameter in the context of heat flux heating, 
whereas \cite{Aubert09} refer to $\Ray_Q^\star$ as being  
controlled only in the limit of vigorous enough convection.
Finally, \cite{Dietrich} introduce $\Ray_Q^\star$ as a control parameter, 
but the parameter they denote as $\Ray_Q^\star$ corresponds to a slightly different definition from that of 
\cite{Chr06}: it involves the heat flux at the outer 
boundary. With this definition, $\Ray_Q^\star$ becomes a control parameter 
entering their governing equations.

The issue of the nature of the $\Ray_Q^\star$ parameter is important, 
since this parameter is used to quantify the injected power 
involved in the power-based scaling laws. Such laws are then re-interpreted 
in the context of natural objects.  
The relation between $\Ray_Q^\star$ and the injected power has first been 
highlighted by \cite{Chr06} in the particular context of convective dynamos, 
driven by an imposed difference of temperature between boundaries, in the 
spherical geometry with a linear radial profile of gravity. 
In this context, they stressed that for vigorous enough flows, 
the injected power is proportional to 
$\Ray_Q^\star\,.$

The configuration in \cite{Chr06} is however one among the numerous 
existing configurations in the literature on convective dynamos. 
Numerous mechanisms for driving convection have indeed been considered. 
Table~1 in \cite{Ku02}, for example, gathers a sample of the thermal or chemical boundary 
conditions implemented in numerical dynamos. Either the temperature, or the 
heat flux, can be fixed at one or both boundaries, and internal heating 
(or secular cooling) can also be implemented through a source (or sink) 
term in the temperature equation. Instead of the isothermal boundary conditions, 
these more complex configurations involving fixed heat flux can be used in an attempt to increase the relevance of numerical models to natural objects. 
They have been investigated in both purely 
convective studies \citep[e.g.][]{Gibbons08}, and in dynamo configurations, as in 
\cite{Saku09} and \cite{Hori12}.

Concerning the domain geometry, more attention has been paid to the spherical 
geometry because of its greater geophysical and astrophysical 
relevance. In this geometry, the radial profile of gravity commonly used by 
geophysicists corresponds to a uniform distribution of mass, and is therefore 
linear \citep[e.g.][]{Chr01}, whereas studies 
motivated by giant planets and stars correspond to a central mass and have thus been performed with a gravity profile 
proportional to $1/r^2$ \citep[e.g.][]{Jones11}. 
In a purely hydrodynamical context, \cite{Gastine15} tested the effect of 
various radial distributions of gravity 
on the boundary layer asymmetry. Nevertheless, the cartesian geometry is 
also interesting, as stressed by several recent studies 
\citep[e.g.][]{Stellmach,Tilgner14, Guervilly15}. 

The issue of the robustness of the relation between the injected power 
and the $\Ray_Q^\star$ parameter introduced by \cite{Chr06} 
is essential. It is important to understand to what extent such a relation
can be used in numerical models and in planetary dynamos, and how it is  
modified by both the geometry and the driving mechanism. 
\cite{Gastine15} shows analytically that the expression derived by \cite{Chr06} 
is valid (up to a geometrical factor) whatever the distance 
to the convection threshold for the particular choice of a gravity profile 
of the form $g \sim r^{-2}$. The case of fixed-flux boundary conditions 
has been examined by \cite{Aubert09} in a study of the palaeo-evolution 
of the geodynamo. Their approach 
is based on the assumption that the total dissipation is
proportional to the difference between the inner- and outer- boundary 
originated mass anomaly fluxes \citep[see][]{Buffett}. This assumption 
however complicates the comparison with the analytical derivations 
of the type of \cite{Chr06}.

The question of the nature (controlled versus measured) of $\Ray_Q^\star$, 
depending on the driving mechanism for convection clearly 
represents a gap in the literature. This paper aims at clarifying 
this issue, in a first part. The second objective of this paper is to 
further investigate the relation between the injected power and 
the flux-based Rayleigh number, in order to stress under which conditions 
the two quantities are proportional one to the other. 
The effect of the geometry and of the thermal heating mechanism 
is addressed. Our study is based on an analytical approach, supported by a 
database of numerical simulations.

\section{Governing equations and control parameters}

We study dynamos in the rotating thermal convection problem. 
The governing equations in the rotating reference frame under the 
Boussinesq approximation can be written in their non-dimensional form as
\begin{equation}
\partial _t \bmu +  (\bmu {\bm \cdot} \bmnabla) \bmu
=
- \bmnabla \pi 
+ {\Pra} \, \nabla^2 \bmu
- 2 \frac{\Pra}{\Ek} \bfe_\varOmega \times \bmu 
+ \Ray \, \Pra  \, \theta \, \bfe_g
+ \left(\bmnabla \times \bmB \right) \times \bmB\, , \,\,\,\,\,\,
\label{eq_NS}
\end{equation} 
\begin{equation}
\partial _t \bmB = \bmnabla \times (\bmu \times \bmB) 
+ \frac{\Pra}{\Pm} \, \nabla^2 \bmB \, ,
\label{eq_B}
\end{equation}
\begin{equation}
\partial_t \theta + (\bmu {\bm \cdot} \bmnabla) \left(\theta+ T_s\right)
= \nabla^2 \theta \, ,
\label{eq_theta}
\end{equation} 
%\ \vskip -17mm
\begin{equation}
\bmnabla {\bm \cdot} \bmu = \bmnabla {\bm \cdot} \bmB = 
0\, ,
\label{eq_div}
\end{equation} 
where $\bmu$ is the velocity field, $\bmB$ the magnetic
field and $\theta$ the deviation from the conductive temperature profile 
$T_s$. In the following, 
the total temperature will be denoted as ${T}$.
The unit vectors $\bfe_\varOmega$ and $\bfe_g$ indicate the direction of the 
rotation axis and of gravity, respectively. They are defined such that ${\bm \varOmega}=\varOmega \, \bfe_\varOmega$ and ${\bm g}=- g \, \bfe_g$. 

The system~(\ref{eq_NS}--\ref{eq_div}) has been written by using $d$ as 
unit of length, $d^2/\kappa$ as unit of time and $\sqrt{\rho \mu}\kappa/d$ as 
unit of magnetic field. It yields the  
non-dimensional parameters $\Ek = \nu / (\varOmega d ^2)$ (Ekman number), 
$\Pra={\nu}/{\kappa}\,$ (Prandtl number) and $\Pm={\nu}/{\eta}\,$ 
(magnetic Prandtl number), where $\nu$ is the 
kinematic viscosity of the fluid, $\kappa=k/(\rho c)$ is the thermal 
diffusivity, $\eta$ is the magnetic diffusivity and 
$\varOmega$ is the rotation rate. We introduce the Rayleigh number 
\begin{equation}
\Ray = \frac{\alpha g \Delta T^\star d^3}{\nu \kappa }\,,
\label{Rayleigh}
\end{equation}
where $\alpha$ is the coefficient of thermal expansion, $g$ 
the gravitational acceleration and $\Delta T^\star$ the difference 
of temperature between both boundaries. Note that in case of an 
imposed flux boundary condition, the temperature must be averaged both in space 
(on the sphere) and in time in order to obtain a unique value for 
$\Delta T^\star$. Dimensional 
quantities are denoted with a star ($^\star$).

Our analysis will be tested against a 
wide database of 184 direct
numerical simulations kindly provided by U. Christensen, corresponding 
to rigid boundaries. Most of them were previously reported in \cite{Chr06}, \cite{Olson06} and \cite{Chr10}. It covers the parameter range $\Ek \in \left[10^{-6},10^{-3}\right]$,  $\Pra \in \left[10^{-1},10^{2}\right]$, $\Pm \in \left[4\times 10^{-2},66.70\right]$ and $\Ray \in \left[10^{5},2.2 \times 10^{9}\right].$

The nature (controlled versus measured) of parameters which measure 
the strength of convection depends on the thermal boundary conditions. 
For imposed temperature at both boundaries 
(differential heating, $\dh$), the unit of temperature is 
$\Delta T^\star$, and the Rayleigh number $\Ray$ defined in 
(\ref{Rayleigh}) is a control parameter. Such is however not the case 
in configurations 
with fixed heat flux (fixed-flux heating, $\ff$). In such configurations, 
either the heat flow $Q^{\star}$ is fixed at both boundaries, in which case 
the temperature is defined up to a constant, or $Q^{\star}$ is fixed 
at the outer 
boundary and the temperature is fixed at the inner boundary. 
For fixed-flux heating, a natural choice of unit of temperature 
is then $\varepsilon^2 Q^{\star}/(\kappa \rho c d)$.  
It involves a factor $\varepsilon$ related to the 
geometry of the domain 
(defined later in the text). It is convenient 
to define a modified Rayleigh number 
\begin{equation}
\Ray_{\varPhi}=\frac{\alpha g  \varepsilon^2 Q^{\star} d^2}{\rho c \nu \kappa^2} \,,
\label{Rayleighphi}
\end{equation}
where $Q^{\star}$ is the time-averaged heat flow across the layer 
(Joules per second).
In fixed-flux heating, this parameter is indeed controlled, 
whereas the classical Rayleigh number is not. 

The Nusselt number $\Nu$ allows to characterize the convective heat transport. 
It is defined as the ratio of the total time-averaged heat flux across the layer $Q^{\star}$ to the ``purely 
diffusive'' heat flow $Q_{\mathrm {cond}}^{\star}$, which corresponds to the heat flux 
which would be measured in the layer if the fluid was at rest with the realized 
$\Delta {T^{\star}}$. This last parameter corresponds to the difference 
between the temperature averaged in time and on both boundaries. Hence, 
$Q_{\mathrm {cond}}^{\star}=\varepsilon^{-2} \kappa \rho c d \Delta T^\star$ and 
\begin{equation}
\Nu=\frac{Q^\star}{\varepsilon^{-2} \kappa \rho c d \Delta T^\star}\,,
\label{Nusselt1}
\end{equation}
that can be re-expressed as 
\begin{equation}
\Nu=\frac{\Ray_{\varPhi}}{\Ray} \,.
\label{Nusselt2}
\end{equation}

Besides, the quantity $Q_{\mathrm{adv}}^\star=Q^\star-Q_{\mathrm {cond}}^\star$ is often used in the 
literature because independent on the vertical/radial coordinate (in the cartesian/spherical geometry). It allows to define the flux-based Rayleigh number as
\begin{equation}
\Ray_Q^\star=\frac{\alpha g Q_{\mathrm{adv}}^\star \varepsilon^2}{\rho c \varOmega^3 d^4}
=\Ray (\Nu-1) \frac{\Ek^3}{\Pra^2}=(\Ray_{\varPhi}-\Ray) \frac{\Ek^3}{\Pra^2}\,.
\label{RastarQ}
\end{equation}
The above expression can be re-expressed as 
\begin{equation}
\Ray_Q^\star \frac{\Nu}{\Nu-1}=\frac{\Ray_{\varPhi}\Ek^3}{\Pra^2}\,.
\label{RastarQ2}
\end{equation}

In both differential and fixed-flux heating, $\Ray_Q^\star$ is not 
a control parameter, since it involves both $\Ray$ and $\Ray_{\varPhi}$, which 
are respectively controlled, depending on the thermal boundary conditions. 
Nevertheless,  according to (\ref{RastarQ2}), in the case of fixed-heat flux 
and if the convection is vigorous enough 
(i.e. $\Nu\gg 1$), $\Ray_Q^\star$ tends to a combination of control parameters
$\Ray_{\varPhi}\Ek^3/\Pra^2$.
The approximation  $\Nu\gg 1$ is very sensible for natural objects (stars 
or planets), but is not well justified for numerical dynamos (in present 
simulations, most dynamos operate at $\Nu < 10$).

\section{Relation between injected power and the flux-based Rayleigh number}

The success of the $\Ray_Q^\star$ parameter relies on its relation with 
the power injected by buoyancy forces, first derived in \cite{Chr06}. 
The injected power $\Power^\star$ (in units Joule per second) in 
its dimensional form is
\begin{equation}
\Power^\star=\iiint \rho \alpha g \theta^\star \bfe_g \, {\bm \cdot} \, \bmu^\star \,\dd V^\star \,,
\end{equation}
which, in non-dimensional form (in units $\rho \kappa^3 d^{-1}$), becomes
\begin{equation}
\left[\begin{array}{l}
\Power^\dh\\[0.2em]
\Power^\ff
\end{array}\right]
\,=\,\left[\begin{array}{l}
\Ray \Pra \\[0.2em]
\Ray_{\varPhi} \Pra
\end{array}\right]\iiint f_g \theta \, \bfe_g \, {\bm \cdot} \, \bmu \,\,\dd V \,,
\label{Pow}
\end{equation}
where $f_g$ is a factor which depends on the geometry and on the radial profile of $g$. 
In the cartesian geometry with uniform gravity, $f_g=1$. In the spherical 
geometry, the gravity 
$g$ involved in the definitions (\ref{Rayleigh}) and 
(\ref{Rayleighphi}) of $\Ray$ and $\Ray_{\varPhi}$ corresponds to the value of gravity at $r_o$. This leads to $f_g=\left(r_o^2/d^2\right) r^{-2}$ for 
$g \sim r^{-2}$, and $f_g=\left(d/r_o\right) r$ for $g \sim r$.
Note that in the above expressions, $\theta$ can equivalently be replaced by 
the total temperature $T$, because the integral over the volume of 
$\left(T_s \, \bfe_g \, {\bm \cdot} \, \bmu\right)$ vanishes.

This section aims at studying how the relation between the injected power 
and the $\Ray_Q^\star$ parameter is affected by the geometry (cartesian versus 
spherical geometry, profile of gravity) and by the thermal boundary conditions (differential versus fixed-flux heating).  

\subsection{Expressions of heat flows in the cartesian and spherical geometries}

In the cartesian configuration that is usually considered, the unit vectors $\bfe_\varOmega$ and $\bfe_g$ are parallel to the $z$-axis. The boundaries are separated by a distance $d$, and located at the planes $z=0$ and $z=1$. The horizontal coordinates are denoted as $x$ and $y$, and vary between $0$ and $\varepsilon^{-1}$, where $\varepsilon=d/\rm{L}$ ($\rm{L}$ being the physical length of the domain in the $x$ and $y$ directions). The gravity is assumed to be uniform in the domain. The conductive temperature profile is solution of $\nabla^2 T_s=0$. The choice of unit of temperature fixes $\,\dd {T_s}/\,\dd z$ to unity, and the constant is chosen such that $ T_s(0)=1$, this leads to ${T_s}=1-z\,.$

The heat flow $Q^{\star}$ across the layer, which is independent on $z$, is the sum of the conductive heat flow $Q_c^{\star}$ and the advective heat flow 
$Q_a^{\star}$, both of these being dependent on $z$. They are defined as
\begin{equation}
Q_c^{\star}(z)= \kappa \rho c \iint - \bmnabla({T_s^{\star}}+\theta^{\star}) \, {\bm \cdot} \, \bfe_g \, \dd x^{\star} \dd y^{\star}, \qquad\quad Q_a^{\star}(z) = \rho c \iint {T^{\star}}\, u_z^{\star}\, \dd x^{\star} \dd y^{\star} \,. \qquad
\end{equation}

Using the above expression of $T_s$, the conductive heat flow becomes
\begin{equation}
\left[\begin{array}{l}
{Q_c^{\star}}^\dh(z)\\[0.2em]
{Q_c^{\star}}^\ff(z)
\end{array}\right]\,=\,
\left[\begin{array}{c}
\kappa \rho c d\Delta { T^{\star}}\\[0.3em]
\varepsilon^2 Q^\star
\end{array}\right]
\left(\varepsilon^{-2} - \iint\frac{\partial \theta}{\partial z} \,\dd x \,\dd y\right)\,
\label{Qccart}
\end{equation}
and the advective heat flow can be re-written as
\begin{equation}
\left[\begin{array}{l}
{Q_a^{\star}}^\dh(z)\\[0.2em]
{Q_a^{\star}}^\ff(z)
\end{array}\right]\,=\,
\left[\begin{array}{c}
\kappa \rho c d\Delta { T^{\star}}\\[0.3em]
\varepsilon^2 Q^\star
\end{array}\right]\iint {T}\, u_z\, \dd x\,\dd y \,.
\label{Qacart}
\end{equation}

In the spherical configuration, the unit vector $\bfe_g$ is radial, and is thus denoted as $\bfe_r$ in following. The boundaries are spherical, and they are located between $r_i$ and $r_o$ ($r_i$ and $r_o$ are dimensional). In our study, the radial profile of gravity is assumed to be either linear, or proportional to $1/r^{2}$. The parameter $d$ corresponds to the thickness of the shell $r_o-r_i$, and the geometrical factor $\varepsilon$ is here defined as $\varepsilon^2=d^2/(4 \pi r_o r_i)=\left(1-\chi\right)^2/\left(4 \pi \chi\right)$, where $\chi=r_i/r_o$. Replacing $\varepsilon^2$ by its definition in (\ref{RastarQ}) yields the expression introduced by \cite{Chr06}
\begin{equation}
\Ray_Q^\star=\frac{1}{4 \pi r_o r_i}\, \frac{\alpha g Q_{\mathrm{adv}}^\star}{\rho c \varOmega^3 d^2}\,. 
\end{equation}
Concerning the conductive temperature profile $T_s$, 
the choice of temperature units fixes 
$r^2\,\dd {T_s}/\,\dd r=-\chi (1-\chi)^{-2}$ and 
we chose the constant such that ${T_s}(r_i/d)=1$. This yields 
$T_s=\chi (1-\chi)^{-2} r^{-1} -\chi (1-\chi)^{-1}$ 
and the corresponding dimensional expressions
\begin{equation}
{{T_s^\star}^\dh}= \frac{\Delta {T^\star} r_i r_o}{d}\frac{1}{r^\star}+\left(1-\Delta {T^\star} \frac{r_o}{d}\right), \qquad\quad
{{T_s^\star}^\ff} = \frac{Q^\star}{4 \pi \kappa \rho c}\frac{1}{r^\star}+ \left(1 - \frac{Q^\star}{4 \pi \kappa \rho c r_i}\right).\qquad
\label{Tsph}
\end{equation}
Using (\ref{Tsph}) allows to re-express the conductive heat flow
\begin{equation}
Q_c^{\star}(r)= \kappa \rho c \iint - \bmnabla({T_s^{\star}}+\theta^{\star}) \, {\bm \cdot} \bfe_r \, r^2 \sin \theta \,\dd\theta\, \dd\phi 
\end{equation}
as
\begin{equation}
\left[\begin{array}{l}
{Q_c^{\star}}^\dh(r)\\[0.2em]
{Q_c^{\star}}^\ff(r)
\end{array}\right]\,=\,
\left[\begin{array}{c}
\kappa \rho c d  \Delta {T^{\star}}\\[0.3em]
\varepsilon^{2} Q^\star
\end{array}\right]\left(\varepsilon^{-2}  - \iint \frac{\partial \theta}{\partial r} {r}^2 \sin \theta \,\dd\theta\, \dd\phi\right),
\label{Qcsph}
\end{equation}
and the advective heat flow
\begin{equation}
Q_a^{\star}(r) = \rho c \iint {T^{\star}}\, u_r^{\star}\, r^2 \sin \theta \,\dd\theta\,\dd\phi \,, 
\end{equation}
 as
\begin{equation}
\left[\begin{array}{l}
{Q_a^{\star}}^\dh(r)\\[0.2em]
{Q_a^{\star}}^\ff(r)
\end{array}\right]\,=\,
\left[\begin{array}{c}
\kappa \rho c d  \Delta {T^{\star}}\\[0.3em]
\varepsilon^{2} Q^\star
\end{array}\right]\iint T\, u_r\, {r}^2 \sin \theta \,\dd\theta\,\dd\phi\,.
\label{Qasph}
\end{equation}
The above expressions of the conductive and advective heat flows will be used 
to re-express the injected power (\ref{Pow}) as a function of the $\Ray_Q^\star$ 
parameter in the two next sections.

\subsection{Differential heating}

For differential heating, the Nusselt number defined in (\ref{Nusselt1}) 
can be re-expressed in the cartesian geometry, using (\ref{Qccart}a) and (\ref{Qacart}a) (Hereafter the labels ``a'' and ``b'' used on matrix equation references refer to the top and bottom row, respectively.), as
\begin{equation}
\Nu=1- \varepsilon^2 \!\iint \frac{\partial \theta}{\partial z} \,\dd x \,\dd y + \varepsilon^2 \!\iint {T}\, u_z\, \dd x \,\dd y \, , 
\label{Nu1}
\end{equation} 
and in the spherical geometry, using (\ref{Qcsph}a) and (\ref{Qasph}a), as
\begin{equation}
\Nu=1- \varepsilon^2 \!\iint \frac{\partial \theta}{\partial r}  r^2 \sin \theta \, \dd \theta \, \dd \phi + \varepsilon^2\! \iint T\, u_r\,  r^2 \sin \theta \,\dd\theta\, \dd \phi \,.
\label{Nu2}
\end{equation}
On time average and in the absence of internal sources or sinks of energy, the above expressions are independent on the radius $r$.\\

In the cartesian geometry, the expression (\ref{Nu1}) allows to rewrite the injected power (\ref{Pow}a) as  
\begin{equation}
\Power^\dh= \Ray \Pra \int \left[\varepsilon^{-2} (\Nu-1) + \iint \frac{\partial \theta}{\partial z} \,\dd x \, \dd y \right] \dd z \,. 
\end{equation}
The integral in the second term vanishes because $\theta$ is zero at both boundaries, which leads to
\begin{equation}
\Power^\dh= \left(\frac{\Pra}{\Ek}\right)^{\!3} \Ray_Q^\star \,\varepsilon^{-2} \,.
\label{P1}
\end{equation}

In the spherical geometry with a radial profile of gravity of the form 
$g \sim r^{-2}$, the injected power (\ref{Pow}a) can be re-expressed as
\begin{equation}
\Power^\dh=\Ray \Pra \frac{r_o^2}{d^2} \iiint \,{T} \, \bfe_r {\bm \cdot} \bmu \,\dd r \sin \theta \, \dd \theta \,\dd \phi  \,.
\label{P_sph}
\end{equation}
Re-writing (\ref{P_sph}) as 
\begin{equation}
\Power^\dh=\Ray \Pra \frac{r_o^2}{d^2} \int \frac{1}{r^2} \left[ \iint {T} \, u_r  r^2 \sin \theta \,\dd \theta \,\dd \phi \right] \dd r\,,
\label{int}
\end{equation}
allows to inject the expression (\ref{Nu2}) of the Nusselt number, which leads to 
\begin{equation}
\Power^\dh=\Ray \Pra \frac{r_o^2}{d^2} \int \frac{1}{r^2} \left[\varepsilon^{-2} (\Nu-1) + \iint \frac{\partial \theta}{\partial r} r^2 \sin \theta \, \dd \theta \, \dd \phi \right] \dd r \,. 
\end{equation}
The second term vanishes because of the boundary conditions, and we finally obtain
\begin{equation} 
\Power^\dh=\left(\frac{\Pra}{\Ek}\right)^{\!3} \frac{4 \pi}{\left(1-\chi\right)^2} \,
\Ray_Q^\star \,.
\label{P2}
\end{equation}
As expected, in the limit of a thin layer 
($\chi\to 1$), this result tends to the cartesian result (\ref{P1}), since 
\begin{equation}
\varepsilon^2 \left[ \left(\frac{\Pra}{\Ek}\right)^{\!3} \frac{4 \pi}{\left(1-\chi\right)^2} \Ray_Q^\star \right] \underset{\chi\to 1}{\longrightarrow} \left(\frac{\Pra}{\Ek}\right)^3 \Ray_Q^\star\,.
\end{equation}

The conclusions at this stage are that, for differential heating, 
there is an exact linear relation between the injected power and 
the $\Ray_Q^\star$ parameter in the cartesian geometry, just like 
in the spherical geometry with a gravity profile proportional to $1/r^2$  
\citep[see also][for this last configuration]{Gastine15}.
The proportionality factor however depends on the geometry. 

Let us now focus on the geometry studied in \cite{Chr06}, that is to say 
the spherical geometry with a linear radial profile of gravity. 
Here we aim at calculating the relation between the injected power 
and $\Ray_Q^\star$ without any approximation. 
In this configuration, the convective power (\ref{Pow}a) can be rewritten as
\begin{equation}
\Power^\dh=\Ray \Pra \frac{d}{r_o} \iiint \,{T} \, \bfe_r {\bm\cdot}\bmu \, r^3 \,\dd r\, \sin \theta \, \dd \theta\,\dd \phi \,. 
\end{equation}

\begin{figure}
\centerline{
\includegraphics*[width=0.7 \textwidth]{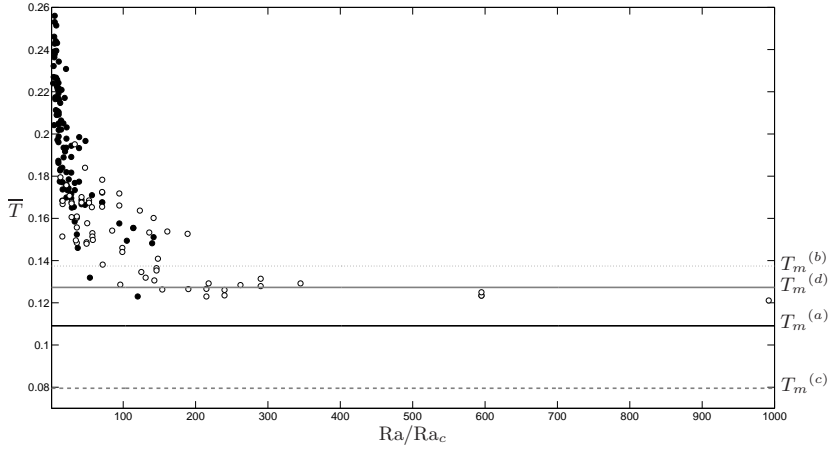}
}
\caption{Representation of the mean temperature $\overline{T}$ as a function of the Rayleigh number, normalized by the critical Rayleigh number at the threshold of convection. Points correspond to the full 184 dynamos database of U. Christensen. Bullets correspond to dipolar dynamos, circles to multipolar ones. Lines correspond to estimations of $\overline{T}$ : ${T_m}^{(a)}$ (solid black), ${T_m}^{(b)}$ (dotted gray), ${T_m}^{(c)}$ (dashed gray) and 
${T_m}^{(d)}$ (solid gray).}
  
\label{Fig1}
\end{figure}

Using expression (\ref{Nu2}) for the Nusselt number yields 
\begin{equation}
\Power^\dh= \Ray \Pra \frac{d}{r_o} \int r \left[\varepsilon^{-2} (\Nu-1) + \iint \frac{\partial \theta}{\partial r} r^2 \sin \theta \,\dd \theta \,\dd \phi \right]\dd r \,,
\end{equation}
which can be re-expressed as
\begin{equation}
\Power^\dh = 2 \pi \chi \frac{1+\chi}{\left(1-\chi\right)^2}\left(\frac{\Pra}{\Ek}\right)^{\!3} \Ray_Q^\star + \Ray \Pra \left(1-\chi\right) \iiint r^3   \frac{\partial \theta}{\partial r} \,\dd r \,\sin \theta \,\dd \theta \,\dd \phi \,. 
\end{equation}
Contrary to what happens in the cartesian geometry and in the spherical 
geometry with $g\sim r^{-2}$, the second term here involves an $r^3$ factor. 
This term therefore does not vanish.  
An integration by part leads to 
\begin{equation}
\Power^\dh = 2 \pi \chi \frac{1+\chi}{\left(1-\chi\right)^2}\left(\frac{\Pra}{\Ek}\right)^{\!3} \Ray_Q^\star - 3 \Ray \Pra  \left(1-\chi\right)  \iiint \,\theta  r^2 \,\dd r\, \sin \theta \, \dd \theta \,\dd \phi \,, 
\end{equation}
and that provides, after some rearrangements, 
\begin{equation}
\Power^\dh =  \left(\frac{\Pra}{\Ek}\right)^3 \Ray_Q^\star \left[2 \pi \chi \frac{1+\chi}{\left(1-\chi\right)^2} - 3 \left(1-\chi\right)  V  \frac{\overline{\theta}}{\Nu-1} \right] ,
\label{P3}
\end{equation}
where the overbar indicates the mean over the volume of the shell, denoted 
as $V$. In the dimensional
form, this yields 
\begin{equation}
{\Power^\star}^\dh= \left(\rho \varOmega^3 d^5\right) \Ray_Q^\star \left[2 \pi \chi \frac{1+\chi}{\left(1-\chi\right)^2} - 3 \left(1-\chi \right)  V  \frac{\overline{\theta}}{\Nu-1}\right].
\label{P3_dim}
\end{equation} 
In this case, the relation between the injected power and $\Ray_Q^\star$ is not 
purely linear. Relation (\ref{P3}) indeed exhibits 
an additional term in the square brackets, which corresponds to the 
relative correction from the linear relation between $\Power$ and $\Ray_Q^\star$. It
is proportional to the mean temperature perturbation over the 
shell, and stems from the 
assumption of a uniform distribution of mass in the spherical geometry. 
An estimation of this term can be made by estimating $\overline{\theta}$.
We assume that in the boundary layers, the heat is purely transported by conduction. This hypothesis is combined to the assumption that the fluid is isothermal in the bulk (which is all the more verified that the convection is vigorous). The temperature in the bulk 
is denoted as ${T_m}$, and corresponds to the mean temperature over the shell $\overline{{T}}$, under the hypothesis of thin boundary layers. This yields 
\begin{equation}
\Bigl(\frac{r_i}{d}\Bigr)^{\!2}\, \frac{{T_i}-{T_m}}{\delta_i} \simeq \Bigl(\frac{r_o}{d}\Bigr)^{\!2}\,\frac{{T_m}-{T_o}}{\delta_o}  \,,
\label{esti}
\end{equation}
where $\delta_i$ and $\delta_o$ correspond to the thickness of the inner and outer boundary layers, respectively. A crude approach consists in assuming that the boundary layers are 
symmetric (i.e. $\delta_i=\delta_o$). The choice ${T_i}=1$ and 
${T_o}=0$ leads to ${T_m}^{(a)}=\chi^{2}/(1+\chi^{2})$, which logically 
tends to $1/2$ as $\chi$ tends to unity. In the literature of 
convection in spherical geometry, alternative physical assumptions have been 
proposed to close the system (\ref{esti}) \citep[e.g.][]{Gastine15}. 
The assumption that thermal boundary layers are marginally stable 
\citep[][]{Vangelov94} provides, in this configuration, a value of the 
mean temperature in the bulk 
of ${T_m}^{(b)}=\chi^{7/4}/(1+\chi^{7/4})$. 
The proposition by \cite{Wu91} that the thermal boundary layers adapt 
their temperature drops and their thicknesses such that their temperature 
scales $\nu \kappa /(g \alpha \delta^3)$ are equal yields 
${T_m}^{(c)}=\chi^{7/3}/(1+\chi^{7/3})$. Finally, very recently, \cite{Gastine15} proposes that the inner and outer boundary layers exhibit the same density of plumes, which leads to ${T_m}^{(d)}=\chi^{11/6}/(1+\chi^{11/6})$. These four estimates can be tested against the numerical database provided by U. Christensen, which corresponds to convective dynamos driven by 
differential heating and a linear radial profile of gravity. The aspect 
ratio is $\chi=0.35$, which provides the estimates ${T_m}^{(a)} \simeq 0.1091$, ${T_m}^{(b)} \simeq 0.1374$, ${T_m}^{(c)} \simeq 0.0795$ and ${T_m}^{(d)} \simeq 0.1273$. Figure \ref{Fig1} shows the mean temperature in 
the shell as a function of the Rayleigh number, normalized by the critical 
Rayleigh number at the threshold of convection. Dynamos exhibiting a dipolar 
magnetic field are distinguished from multipolar dynamos. We observe that 
the mean temperature tends to a constant value, as the Rayleigh number 
increases. Both estimates ${T_m}^{(a)}$ and ${T_m}^{(d)}$ are remarkably well met by the multipolar dynamos associated to the most vigorous convection. 
We can notice that the simplest hypothesis (symmetric boundary layers) provides 
a very good agreement to the numerical database, which is not significantly 
improved by considering more elaborate assumptions.

Let us note that replacing $\overline{T}$ by the estimated value
${T_m}^{(a)}$, and $\overline{T_s}$ by
\begin{equation} 
\overline{{T_s}}= \frac{\chi \left(-2\chi^2+\chi+1\right)}{2 \left(1-\chi^3\right)}\,,
\end{equation}
in (\ref{P3}) yields
\begin{equation}
{\Power}^\dh= \left(\frac{\Pra}{\Ek}\right)^{\!3} \Ray_Q^\star \left[2 \pi \chi \frac{1+\chi}{\left(1-\chi\right)^2} + \frac{2 \pi f(\chi)}{\Nu-1} \right] \qquad {\rm with} \qquad f(\chi)= \frac{\chi (1+\chi)}{(1+\chi^2)} \,, \qquad 
\end{equation}
and that in the limit $\chi\to 1$,
\begin{equation}
\varepsilon^2 \Power^\dh  
\underset{\chi\to 1}{\longrightarrow} \left(\frac{\Pra}{\Ek}\right)^{\!3} \Ray_Q^\star  \,, 
\end{equation}
which is consistent with the expression (\ref{P1}) derived in the cartesian 
geometry. 

\begin{figure}
\centerline{
\includegraphics*[width=0.70 \textwidth]{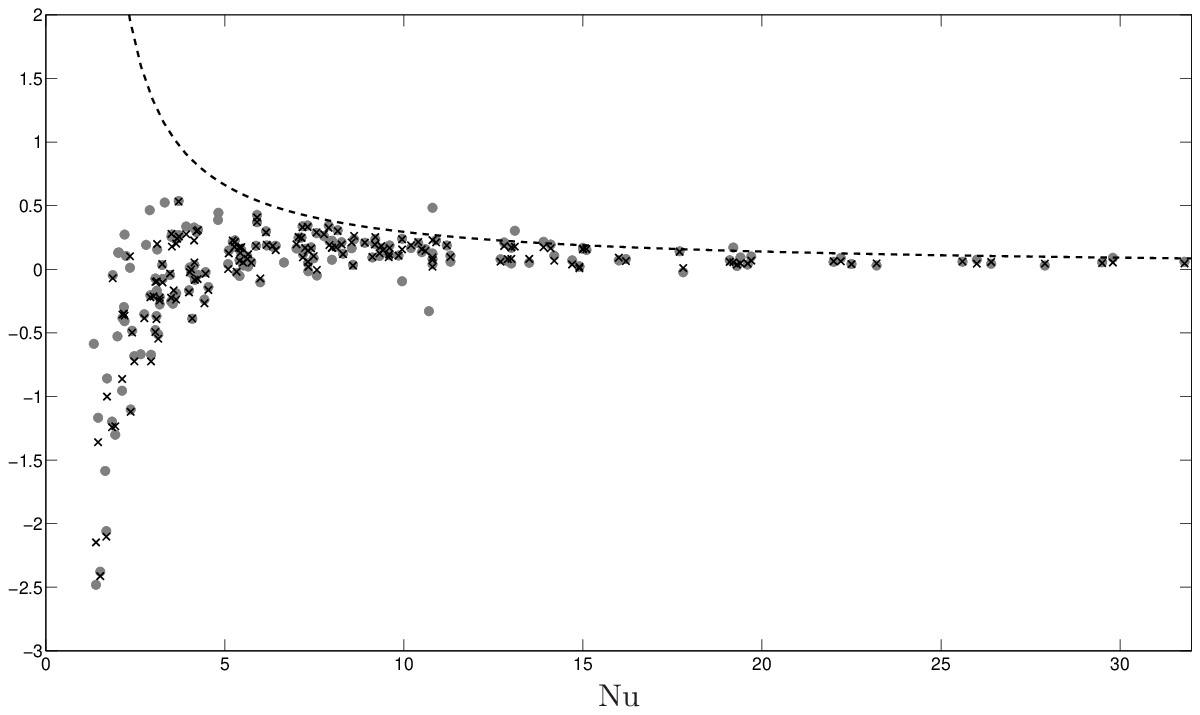} 
}
\caption{Representation of the relative correction term in (\ref{P3}), in the form $\bigl[\Power^\dh-2\pi \chi (1+\chi)(1-\chi)^{-2}\Ray_{Q}^\star\bigr]\big/\Ray_{Q}^\star$ (gray bullets) and $-3 \left(1-\chi\right) V \overline{\theta}\left(\Nu-1\right)^{-1}$ (black crosses) as a function of the Nusselt number $\Nu$, using the numerical database provided by U. Christensen ($\Power^\dh$ is here in units $\rho \varOmega^3 d^5$). 
 The dashed line corresponds to the analytic function $f(\Nu)=-3 \left(1-\chi\right) V ({T_m}^{(a)} - \overline{{T_s}})\left(\Nu-1\right)^{-1}$; as expected, data tend asymptotically to this function for vigorous enough convection. 
This figure validates relation (\ref{P3}).}
 
\label{Fig2}
\end{figure}

Figure \ref{Fig2} allows to test relation (\ref{P3}). It represents 
the relative correction term in (\ref{P3}), in the form 
$$\frac{\Power^\dh-2\pi \chi (1+\chi)(1-\chi)^{-2}\Ray_{Q}^\star}{\Ray_{Q}^\star}$$ and 
$$-3 \left(1-\chi\right) V \frac{\overline{\theta}}{\Nu-1}\,,$$
using the present numerical database ($\Power^\dh$ is here in units $\rho \varOmega^3 d^5$).
The good agreement between both parts validates relation (\ref{P3}).

\begin{figure}
\centerline{
\includegraphics*[width=1. \textwidth]{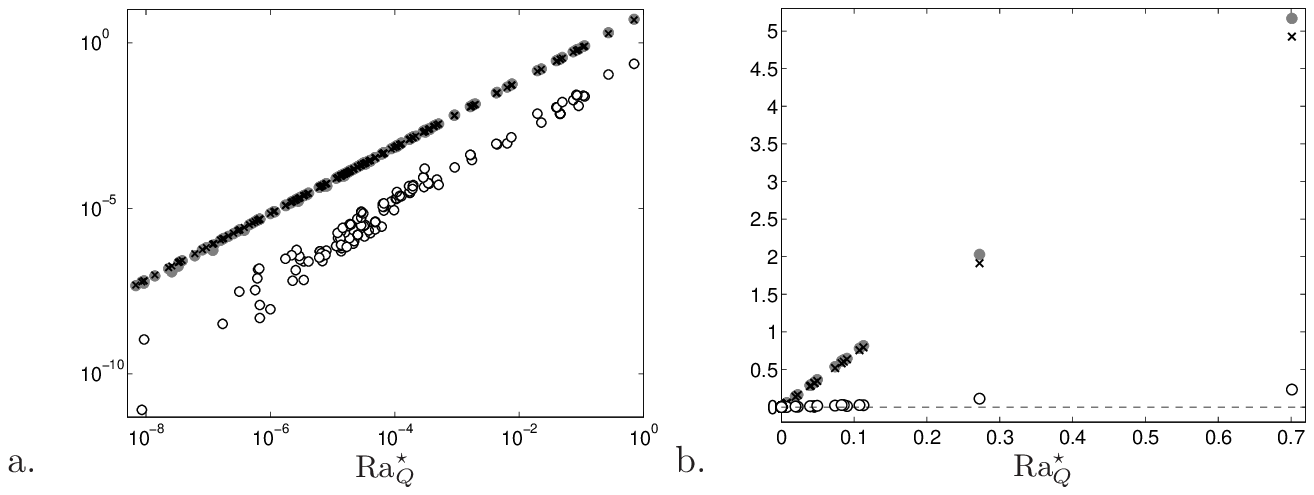} 
}
\caption{The three terms of (\ref{P3_dim}) 
in the form $\Power^\dh$ (in units $\rho \varOmega^3 d^5$, gray bullets), 
$2\pi \chi (1+\chi)(1-\chi)^{-2}\Ray_{Q}^\star$ (black crosses) and 
the absolute correction term $-3 \left(1-\chi\right) V \overline{\theta}\left(\Nu-1\right)^{-1} \Ray_{Q}^\star$ 
(black circles), as a function of the flux-based Rayleigh number, using the 
numerical database provided by U. Christensen. 
Panel (a) corresponds to a log-log representation, and panel (b) to a linear one. }
\label{Fig3}
\end{figure}

The first term on the right-hand side of (\ref{P3_dim}) 
corresponds to the 
expression derived by \cite{Chr06}. The second term, that we have just 
further investigated above, has been neglected in their study. 
This is equivalent to neglecting the contribution of the gradient of the 
temperature perturbation in the conductive heat flow 
(\ref{Qcsph}a), compared to the gradient of the purely conductive profile 
${T_s}$. Such an approximation is of course very sensible near the 
threshold of convection. At the threshold, this term indeed 
vanishes. In order to ponder on its reliability as convection 
becomes more vigorous, we represented in figure \ref{Fig3} the three terms of (\ref{P3_dim}) in the form $\Power^\dh$ (in units $\rho \varOmega^3 d^5$, gray bullets), $2\pi \chi (1+\chi)(1-\chi)^{-2}\Ray_{Q}^\star$ (black crosses) and 
the absolute correction term $-3 \left(1-\chi\right) V \overline{\theta}\left(\Nu-1\right)^{-1} \Ray_{Q}^\star$ (black circles), 
as a function of the flux-based Rayleigh number.
The first panel corresponds to a log-log representation \citep[as in][]{Chr06}, 
whereas the second one is linear. The figure based on logarithmic scales 
shows how the approximation consisting in neglecting the second term on the 
right-hand-side of (\ref{P3_dim}) is sensible for the present database. 
Nevertheless, the linear representation indicates  
that in the numerical database, as the 
$\Ray_{Q}^\star$ parameter becomes larger, this term increases. 
In planets and stars, however,
$\Ray_{Q}^\star\ll 1$  and $\Nu\gg 1$. For example, in the Earth's core, 
$\Ray_{Q}^\star \sim 10^{-13}$ and the Nusselt number based on the superadiabatic 
temperature gradient is about $10^6$ \citep[e.g.][]{Olson06,Chr06}. This
yields a negligible absolute correction term. 
The absolute correction term is also small in most existing 
numerical dynamos, with $\Ray_{Q}^\star \in [10^{-8}, \, 1]$ and 
$\Nu \in [1,\, 32]\, .$
The limit of the linear relation 
between the injected power and $\Ray_{Q}^\star$ in this configuration, however,
should not be ignored by numericists.
Rewriting the absolute correction term as $-3 \left(1-\chi\right) V \, \overline{\theta} 
\, \Ray \, \Ek^3/\Pra^2$ using (\ref{RastarQ}) 
reveals that trying to increase $\Nu$ 
(which is underestimated by a factor $\sim 10^5$ in existing simulations) through 
an increase of $\Ray$ at fixed $\Ek$, will necessarily yield an increase of the 
correction term. However, as $\Ek$ can be decreased 
(with increasing computational resources), 
the Rayleigh number will have to increase as $\Ray_c \sim \Ek^{-4/3}$. 
For fixed super-criticality, the correction term will thus scale as $\Ek^{5/3}$. 
Thus an increase in $\Ray/\Ray_c$ can be achieved, while the correction term 
remains small. This is necessary to achieve a limit relevant to 
planetary cores.

This restriction neither exists in 
the spherical geometry with $g$ proportional to $1/r^2$, nor 
in the cartesian geometry. 
As the original power-based scaling law for the magnetic field strength 
mainly reflects the statistical balance between injected power and 
dissipation \citep{Oruba}, our results allow to understand why the scaling law 
which relates the magnetic field strength to $\Ray_Q^\star$ 
derived by \cite{Chr06} in the spherical geometry with 
$g\sim r$ is also verified in the planar convective-driven 
dynamos studied by \cite{Tilgner14} (see figure 1a therein). Besides,
this highlighting of the role played by the 
gravity profile in the expression of the injected power corroborates 
the observation made by \cite{Raynaud14} that the mass distribution 
has a strong influence on the fluid flow and thus on the dynamo generated 
magnetic field.

\subsection{Fixed-flux at the CMB}

We focus here on a configuration where the temperature is fixed at 
the inner boundary (ICB), which would correspond to the solidification 
temperature of iron at the pressure of the ICB, and the heat flux is fixed 
at the outer boundary (CMB), which is equivalent to considering that the mantle 
controls the heat flux out of the core. 

The imposed heat flow $Q^\star$ is carried by the background temperature profile ${T_s}$. As a consequence, the Nusselt number reduces to $\Nu=1 /\Delta {T}$, and the confrontation of the expressions of ${Q_c^\star}^\ff$ and ${Q_a^\star}^\ff$ in the cartesian geometry  (see (\ref{Qccart}b) and (\ref{Qacart}b)) leads to
\begin{equation}
\iint \frac{\partial \theta}{\partial z} \,\dd x \, \dd y = \iint {T}\, u_z\,\dd x \,\dd y \,,
\label{Nucart}
\end{equation}
and (\ref{Qcsph}b) and (\ref{Qasph}b) provide, in the spherical geometry, 
\begin{equation}
\iint \frac{\partial \theta}{\partial r} r^2 \sin \theta \, \dd \theta \, \dd \phi = \iint {T} \, u_r  r^2 \sin \theta  \,\dd \theta \,\dd \phi  \,.
\label{Nusph}
\end{equation}
In the cartesian geometry, (\ref{Nucart}) allows to rewrite the injected power (\ref{Pow}b) as
\begin{equation}
\Power^\ff= \Ray_{\varPhi} \Pra \int \left[\iint \frac{\partial \theta}{\partial z} \,\dd x\,\dd y \right] \dd z \,,  
\end{equation}
and replacing $\theta$ by ${T}- {T_s}$ provides
\begin{equation}
\Power^\ff= \Ray_{\varPhi} \Pra  \varepsilon^{-2}\left(-\Delta {T} + 1 \right),  
\end{equation}
which can be rewritten as
\begin{equation}
\Power^\ff=  \left(\frac{\Pra}{\Ek}\right)^{\!3} \Ray_Q^\star \varepsilon^{-2} \,.
\label{P4}
\end{equation}
This expression is thus identical to (\ref{P1}), derived in the differential heating configuration.

In the spherical geometry with $g \sim r^{-2}$, (\ref{Nusph}) allows to rewrite 
the injected power (\ref{int}) as
\begin{equation}
\Power^\ff= \Ray_\varPhi \Pra \,\frac{1}{\left(1-\chi \right)^2} \iiint \frac{\partial \theta}{\partial r}   \,\dd r \,\sin \theta \,\dd \theta \,\dd \phi \,. 
\end{equation}
Introducing the notation $\langle\cdots\rangle=1/(4\pi)\iint \cdots\, \sin \theta \,\dd \theta \,\dd\phi$ yields
\begin{equation}
\Power^\ff= 4 \pi \Ray_{\varPhi} \Pra \,\frac{1}{\left(1-\chi \right)^2}
\bigl({\langle\theta\rangle}_o-{\langle\theta\rangle}_i\bigr), 
\end{equation}
and replacing $\langle\theta\rangle$ by $\langle{T}\rangle-{T_s}$ leads to
\begin{equation}
\Power^\ff= 4 \pi \Ray_{\varPhi} \Pra \,\frac{1}{(1-\chi)^2}\left(
- \Delta {T} + 1\right) \,. 
\end{equation}
We finally obtain
\begin{equation}
\Power^\ff=\left(\frac{\Pra}{\Ek}\right)^{\!3} \frac{4 \pi}{\left(1-\chi\right)^2}\, \Ray_Q^\star  \,.
\label{P5}
\end{equation}

At this stage, the expressions (\ref{P4}) and (\ref{P5}) obtained for 
fixed-flux heating are thus the same as those derived in the differential heating 
configuration (see (\ref{P1}) and (\ref{P2})). 

In the spherical geometry with $g \sim r$, using (\ref{Nusph}) allows to rewrite the injected power (\ref{Pow}b) as 
\begin{equation}
\Power^\ff= \Ray_\varPhi \Pra \frac{d}{r_o} \iiint \frac{\partial \theta}{\partial r}  r^3 \,\dd r \sin \theta  \,\dd \theta \,\dd \phi \,,
\end{equation}
which becomes, after an integration by part
\begin{equation}
\Power^\ff= 4 \pi \Ray_{\varPhi} \Pra \,\frac{d}{r_o} \left[
\Bigl[{\langle\theta\rangle} r^3\Bigr]_{r_i/d}^{r_o/d}
-3 \int r^2 {\langle\theta\rangle} \,\dd r \right].
\label{int2}
\end{equation}
As the temperature at the inner boundary is imposed ($\theta_i=0$), we obtain
\begin{equation}
\Power^\ff= 4 \pi \Ray_{\varPhi} \Pra\, \frac{d}{r_o} \left[
\frac{\langle\theta\rangle_o}{(1-\chi)^3}
-3 \int r^2 {\langle\theta\rangle} \,\dd r \right],
\end{equation}
and replacing $\langle\theta\rangle_o$ by $\langle{T}\rangle_o-\langle{T}\rangle_i+{T_i}-{T_{o}}$ yields
\begin{equation}
\Power^\ff= 4 \pi \Ray_{\varPhi} \Pra \,\frac{d}{r_o} \left[
\frac{1}{(1-\chi)^3}\left(-\Delta {T} +1\right)
-3 \int r^2 {\langle\theta\rangle} \,\dd r\right], 
\label{P6_int}
\end{equation}
and finally
\begin{equation}
\Power^\ff= \left(\frac{\Pra}{\Ek}\right)^{\!3} \Ray_Q^\star 
\left[ 
\frac{4 \pi}{\left(1-\chi\right)^2}  
- 3 (1-\chi) V \frac{\Nu}{\Nu-1}\overline{\theta}  \right].
\label{P6}
\end{equation}
It is interesting to note that as for differential heating, this geometry 
exhibits a supplementary term involving the mean perturbation temperature. 
The correction terms in 
(\ref{P3}) and (\ref{P6}) differ through a $\Nu$ factor, which corresponds 
to the ratio of the two factors in (\ref{Pow}a,b). In the limit of vigorous 
convection, $\overline{\theta}^\dh$ is $\mathrm{O}(1)$ whereas 
$\overline{\theta}^\ff$ is approximately proportional to $1/\Nu$. The correction 
terms in (\ref{P3}) and (\ref{P6}) thus tend to a unique expression, which is 
consistent with the idea that when convection is very vigorous, the effect of
different thermal boundary conditions vanishes \citep[e.g.][]{John09}.
The same restriction for the linear relation between 
the injected power and the flux-based Rayleigh number thus also applies 
to the fixed-flux configuration.  

\subsection{Chemical convection with fixed flux at the ICB}

The most general cases can involve internal volumetric 
sources or sinks of energy. Let us focus on the configuration of
 an imposed uniform heat flux  $Q_i^\star$ at the ICB, zero flux at the CMB 
and homogeneous volumetric sinks \citep[to mimic chemical convection with fixed flux at the ICB, 
as introduced in][]{Ku02}. 
The sink term provides a modification of the conductive temperature profile ${T_s}$,  so that the heat equation (\ref{eq_theta}) remains unchanged. 
In this case, the unit of temperature 
$\mathscr{C}\varepsilon^2 Q_i^{\star}/(\kappa \rho c d)$, 
with $\mathscr{C}=(2+\chi)/(2(1+\chi+\chi^2))$ a geometric factor, 
is chosen such that $\Delta {T_s}=1$. This yields 
\begin{equation}
{T_s}=\frac{\chi}{2+\chi}\left[\frac{-3}{(1-\chi)^2}+r^2+\frac{2}{r (1-\chi)^3}\right],
\end{equation}
where ${T_s}(r_i/d)$ has been set to unity. 
As the time-averaged 
total heat flow $Q^\star$ here depends on radius, the Nusselt number defined in 
(\ref{Nusselt1}) also depends on radius. We therefore define a reference
Nusselt number $\Nu^\dag$ as
\begin{equation}
\Nu^\dag=\frac{\Delta T_s^\star}{\Delta T^\star}=\frac{1}{\Delta T} \,.
\end{equation}
Now, introducing 
\begin{equation}
\Ray_{\varPhi}^\dag=\frac{\alpha g  \varepsilon^2 \mathscr{C} Q_i^{\star} d^2}{\rho c \nu \kappa^2} \, ,
\label{Rayphi2}
\end{equation}
yields $\Nu^\dag={\Ray_{\varPhi}^\dag}/{\Ray}$. Note that 
as $\chi$ tends to unity, 
$\mathscr{C}$ tends to $1/2$. In this limit, the comparison of $\Ray_{\varPhi}$ and $\Ray_{\varPhi}^\dag$ relies on the substitution of the time-averaged total heat flow $Q^\star$ 
(which is a constant in the absence of energy sink) by its volume average (equal to $Q_i^{\star}/2$) in the presence of energy sink. Because of the homogeneous Neumann boundary conditions, the temperature perturbation 
is defined up to an arbitrary constant. If this constant is chosen such that $\langle\theta\rangle_i=0$, 
the expressions (\ref{P5}) and (\ref{P6}) for the injected power, 
derived in the configuration of fixed-flux heating with no volumetric source, 
are recovered in the presence of a sink of energy, provided ${\Ray_Q}^\star$ 
is replaced by
\begin{equation}
(\Ray_{\varPhi}^\dag-\Ray) \frac{\Ek^3}{\Pra^2}= \frac{\alpha g \kappa \left(\Delta {T_s}^\star-\Delta {T}^\star\right)}{\varOmega^3 d^3}\,.
\end{equation}
The conclusions are thus identical to those derived in 
the case of no internal energy sources/sinks.

\section{Conclusions}

In this paper, we first focused on the $\Ray_Q^\star$ parameter 
which plays an essential role in the power-based scaling laws in dynamo. 
We highlighted that for vigorous enough convection (i.e. $\Nu\gg 1$), the 
$\Ray_Q^\star$ parameter tends to a combination of parameters which are 
controlled in the case of fixed-heat flux. However sensible for natural objects, 
this approximation is not straightforward for numerical dynamos in present 
databases. This clarifies the contradictions in the literature concerning the nature 
(controlled versus measured) of this parameter in convective dynamos driven 
by a fixed-heat flux. 

In a second part, we investigated the issue of the robustness of the original 
relation between the power $\Power$ injected by buoyancy forces and the $\Ray_Q^\star$ 
parameter with the geometry and the thermal heating mechanism. This is an important question since this robustness 
is mandatory for the extrapolations of the original power-based scaling 
laws to other configurations to be relevant. 
We have shown analytically that in the cartesian geometry as in the spherical 
geometry exhibiting a radial profile of gravity of the form $1/r^2$, for both 
differential and fixed-flux heating, the linear relation between the injected power 
and $\Ray_Q^\star$ is robust. Only the proportionality factor is modified 
by the geometry. 
The spherical geometry with a linear radial profile of 
gravity is however more complicated, since the relation between $\Power$ and 
$\Ray_Q^\star$ involves an 
additional term proportional to the mean perturbation temperature. 
In the differential heating configuration, we have highlighted and pondered on 
this term by using a numerical database of dynamos. We have shown that it could be 
estimated in the limit of vigorous convection through simple assumptions.  
We have also stressed that the linear approximation 
between $\Power$ and $\Ray_Q^\star$ is relevant to natural dynamos and in 
most existing numerical dynamos. 

Its validity however places an upper-bound on 
the strength of the driving which can be envisaged in a fixed 
Ekman number simulation. An increase of the Rayleigh number 
indeed yields deviations (in terms of absolute correction) 
from the linear relation between $\Power$ and $\Ray_Q^\star$.
The effect of the heating mode on the 
relation between $\Power$ and $\Ray_Q^\star$ is found to be small.
 
To summarize, in convective dynamos driven by a fixed-heat flux, 
the $\Ray_{Q}^\star$ parameter 
becomes controlled provided the Nusselt number is large enough ($\Nu\gg 1$). 
However, in the geometry commonly studied by geophysicists 
(spherical with uniform mass distribution), an increase of the Rayleigh number 
in numerical simulations at fixed Ekman number
would yield a deviation of $\Power$ from $\Ray_{Q}^\star$.
The parameter range for which the power injected by buoyancy 
forces is controlled is thus limited. 
In the quest for predictive scaling laws, an alternative approach to power-based scaling laws consists in expressing forces balances based on the distance to the onset of dynamo action 
\citep[see][]{Petre01,Oruba}.

\section*{Acknowledgments} 
The author wants gratefully thank Uli Christensen 
for sharing his numerical database, and Emmanuel Dormy for useful discussions.

\bibliographystyle{gGAF}
\bibliography{Oruba_2016_Arxiv}

\newcommand{\noopsort}[1]{} \newcommand{\printfirst}[2]{#1}
  \newcommand{\singleletter}[1]{#1} \newcommand{\switchargs}[2]{#2#1}
\begin{thebibliography}{25}
\providecommand{\natexlab}[1]{#1}

\bibitem[\protect\citeauthoryear{Aubert {\itshape{et~al.}}}{2009}]{Aubert09}
Aubert, J., Labrosse, S. and Poitou, C., {Modelling the palaeo-evolution of the
  geodynamo}. {\itshape Geophys. J. Int.}, 2009, \textbf{179}, 1414–1428.

\bibitem[\protect\citeauthoryear{Buffett {\itshape{et~al.}}}{1996}]{Buffett}
Buffett, B.A., Huppert, H., Lister, J. and Woods, A., {On the thermal evolution
  of the Earth's core}. {\itshape J. geophys. Res.}, 1996, \textbf{101(B4)},
  7989--8006.

\bibitem[\protect\citeauthoryear{Christensen}{2010}]{Chr10}
Christensen, U., {Dynamo Scaling Laws and Applications to the Planets}.
  {\itshape Space Sci. Rev.}, 2010, \textbf{152}, 565--590.

\bibitem[\protect\citeauthoryear{Christensen and Aubert}{2006}]{Chr06}
Christensen, U. and Aubert, J., {Scaling properties of convection-driven
  dynamos in rotating spherical shells and application to planetary magnetic
  fields}. {\itshape Geophys. J. Int.}, 2006, \textbf{166}, 97--114.

\bibitem[\protect\citeauthoryear{Christensen {\itshape{et~al.}}}{2001}]{Chr01}
Christensen, U., Aubert, J., Cardin, P., Dormy, E., Gibbons, S., Glatzmaier,
  G., Grote, E., Honkura, Y., Jones, C., Konoh, M., Matsushima, M., Sakuraba,
  A., Takahashi, F., Tilgner, A., Wicht, J. and Zhang, K., {A numerical dynamo
  benchmark}. {\itshape Phys. Earth Planet. Int.}, 2001, \textbf{128}, 25--34.

\bibitem[\protect\citeauthoryear{Davidson}{2013}]{Davidson13}
Davidson, P., {Scaling laws for planetary dynamos}. {\itshape Geophys. J.
  Int.}, 2013, \textbf{195}, 67--74.

\bibitem[\protect\citeauthoryear{Dietrich {\itshape{et~al.}}}{2013}]{Dietrich}
Dietrich, W., Schmitt, D. and Wicht, J., {Hemispherical Parker waves driven by
  thermal shear in planetary dynamos}. {\itshape Europhys. Lett.}, 2013,
  \textbf{104}.

\bibitem[\protect\citeauthoryear{Gastine {\itshape{et~al.}}}{2015}]{Gastine15}
Gastine, T., Wicht, J. and Aurnou, J.M., {Turbulent Rayleigh-Benard convection
  in spherical shells}. {\itshape J. Fluid. Mech.}, 2015, \textbf{778},
  721--764.

\bibitem[\protect\citeauthoryear{Gibbons {\itshape{et~al.}}}{2008}]{Gibbons08}
Gibbons, S., Gubbins, D. and Zhang, K., {Convection in rotating spherical fluid
  shells with inhomogeneous heat flux at the outer boundary}. {\itshape
  Geophys. Astrophys. Fluid Dyn.}, 2008, \textbf{101:5}, 347--370.

\bibitem[\protect\citeauthoryear{Guervilly
  {\itshape{et~al.}}}{2015}]{Guervilly15}
Guervilly, C., Hughes, D.W. and Jones, C.A., {Generation of magnetic fields by
  large-scale vortices in rotating convection}. {\itshape Phys. Rev. E}, 2015,
  \textbf{91}, 041001.

\bibitem[\protect\citeauthoryear{Hori {\itshape{et~al.}}}{2012}]{Hori12}
Hori, K., Wicht, J. and Christensen, U., {The influence of thermo-compositional
  boundary conditions on convection and dynamos in a rotating spherical shell}.
  {\itshape Phys. Earth Planet. Int.}, 2012, \textbf{196-197}, 32--48.

\bibitem[\protect\citeauthoryear{Johnston and Doering}{2009}]{John09}
Johnston, H. and Doering, C., {Comparison of turbulent thermal convection
  between conditions of constant temperature and constant flux}. {\itshape
  Phys. Rev. Lett.}, 2009, \textbf{102}, 064501.

\bibitem[\protect\citeauthoryear{Jones}{2011}]{JonesRev}
Jones, C.A., {Planetary magnetic fields and fluid dynamos}. {\itshape Annu.
  Rev. Fluid Mech.}, 2011, \textbf{43}, 583–614.

\bibitem[\protect\citeauthoryear{Jones {\itshape{et~al.}}}{2011}]{Jones11}
Jones, C., Boronski, P., Brun, A., Glatzmaier, G., Gastine, T., Miesch, M. and
  Wicht, J., {Anelastic convection-driven dynamo benchmarks}. {\itshape
  Icarus}, 2011, \textbf{216}, 120--135.

\bibitem[\protect\citeauthoryear{Kutzner and Christensen}{2002}]{Ku02}
Kutzner, C. and Christensen, U., {From stable dipolar towards reversing
  numerical dynamos}. {\itshape Phys. Earth Planet. Int.}, 2002, \textbf{131},
  29--45.

\bibitem[\protect\citeauthoryear{Olson and Christensen}{2006}]{Olson06}
Olson, P. and Christensen, U.C., {Dipole moment scaling for convection-driven
  planetary dynamos}. {\itshape Earth Planet. Sci. Lett.}, 2006, \textbf{250},
  561--571.

\bibitem[\protect\citeauthoryear{Oruba and Dormy}{2014}]{Oruba}
Oruba, L. and Dormy, E., {Predictive scaling laws for spherical rotating
  dynamos}. {\itshape Geophys. J. Int.}, 2014, \textbf{198}, 828--847.

\bibitem[\protect\citeauthoryear{Petrelis and Fauve}{2001}]{Petre01}
Petrelis, F. and Fauve, S., {Saturation of the magnetic field above the dynamo
  threshold}. {\itshape Eur. Phys. J. B}, 2001, \textbf{22}, 273--276.

\bibitem[\protect\citeauthoryear{Raynaud {\itshape{et~al.}}}{2014}]{Raynaud14}
Raynaud, R., Petitdemange, L. and Dormy, E., {Influence of the mass
  distribution on the magnetic field topology}. {\itshape Astron. Astrophys.},
  2014, \textbf{567}, A107.

\bibitem[\protect\citeauthoryear{Sakuraba and Roberts}{2009}]{Saku09}
Sakuraba, A. and Roberts, P., {Generation of a strong magnetic field using
  uniform heat flux at the surface of the core}. {\itshape Nature Geoscience},
  2009, \textbf{2}, 802--805.

\bibitem[\protect\citeauthoryear{Stellmach and Hansen}{2004}]{Stellmach}
Stellmach, S. and Hansen, U., {Cartesian convection driven dynamos at low Ekman
  number}. {\itshape Phys. Rev.}, 2004, \textbf{70}, 056312.

\bibitem[\protect\citeauthoryear{Stelzer and Jackson}{2013}]{Stelzer13}
Stelzer, Z. and Jackson, A., {Extracting scaling laws from numerical dynamo
  models}. {\itshape Geophys. J. Int.}, 2013, \textbf{193}, 1265--1276.

\bibitem[\protect\citeauthoryear{Tilgner}{2014}]{Tilgner14}
Tilgner, A., {Magnetic energy dissipation and mean magnetic field generation in
  planar convection-driven dynamos}. {\itshape Phys. Rev.}, 2014,
  \textbf{90(1)}, 013004.

\bibitem[\protect\citeauthoryear{Vangelov and Jarvis}{1994}]{Vangelov94}
Vangelov, V.I. and Jarvis, G.T., {Geometrical effects of curvature in
  axisymmetric spherical models of mantle convection}. {\itshape J. Geophys.
  Res.}, 1994, \textbf{99}, 9345--9358.

\bibitem[\protect\citeauthoryear{Wu and Libchaber}{1991}]{Wu91}
Wu, X.Z. and Libchaber, A., {Non-Boussinesq effects in free thermal
  convection}. {\itshape Phys. Rev.}, 1991, \textbf{43}, 2833--2839.

\end{thebibliography}

\end{document}